\documentclass[prl,twocolumn,amsmath,amssymb,amsfonts,final]{revtex4}

\usepackage{graphicx}
\usepackage{color}

\newif\iffigs
\figstrue

\begin{document}

\title{Experimental multiparticle entanglement dynamics induced by decoherence}

\author{ J. T. Barreiro,$^{1}$ P. Schindler,$^{1}$ O. G\"uhne,$^{2,3}$
  T. Monz,$^{1}$\\ M. Chwalla,$^{1}$ C. F. Roos,$^{1,2}$ M. Hennrich,$^{1}$ and
  R. Blatt$^{1,2}$}

\affiliation{$^{1}$Institut f\"ur Experimentalphysik, Universit\"at Innsbruck,
  Technikerstr. 25, 6020 Innsbruck, Austria\\ 
$^{2}$Institut f\"ur Quantenoptik und Quanteninformation, \\
\"Osterreichische Akademie der Wissenschaften, Technikerstr. 21A, 6020 Innsbruck,
  Austria\\
$^{3}$Institut f\"ur Theoretische Physik,
  Universit\"at Innsbruck, Technikerstr. 25, 6020 Innsbruck,
  Austria}

\begin{abstract}
Multiparticle entanglement leads to richer correlations than two-particle
entanglement and gives rise to striking contradictions with local
realism~\cite{horne-ajp-58-1131}, inequivalent classes of
entanglement~\cite{dur-pra-62-062314}, and applications such as one-way or
topological quantum computing~\cite{raussendorf-prl-86-5188,nayak-rmp-80-1083}.
When exposed to decohering or dissipative environments, multiparticle
entanglement yields subtle dynamical features and access to new classes of
states and applications.  Here, using a string of trapped ions, we
experimentally characterize the dynamics of entanglement of a multiparticle
state under the influence of decoherence.  By embedding an entangled state of
four qubits in a decohering environment (via spontaneous decay), we observe a
rich dynamics crossing distinctive domains: Bell-inequality violation,
entanglement superactivation, bound entanglement, and full separability.  We
also develop new theoretical tools for characterizing entanglement in quantum
states.  Our techniques to control the environment can be used to enable novel
quantum-computation, state-engineering, and simulation paradigms based on
dissipation and
decoherence~\cite{diehl-natphys-4-878,verstraete-nphys-5-633,weimer-nphys-6-382}.

\end{abstract}

\maketitle 

When exposed to an environment, bipartite entanglement already shows subtle
dynamical features, e.g., finite-time
disentanglement~\cite{yu-science-323-598,almeida-science-316-579}.  In a
multipartite setting, decoherence and dissipation enable novel quantum
applications~\cite{diehl-natphys-4-878,verstraete-nphys-5-633,weimer-nphys-6-382},
induce even more interesting dynamics due to the different classes of states,
and can decrease the number of particles genuinely entangled --an effect
recently observed~\cite{papp-science-324-764}.  However, decoherence can also
influence many other state properties useful for quantum information
processing, such as distillability and the entanglement between subsystems.  A
state is distillable if, using \emph{local} operations and classical
communication, one can extract the maximally entangled states required by
several quantum communication protocols such as dense
coding~\cite{bennett-prl-69-2881} and teleportation~\cite{bennett-prl-70-1895}.

An environment can also drive a multiparticle entangled and distillable state
into the undistillable but still entangled
domain~\cite{aolita-prl-100-080501,borras-pra-79-022108}, called bound
entangled (BE)~\cite{horodecki-prl-80-5239}.  This class of states is expected
to appear in many body systems~\cite{toth-prl-99-250405}, and despite of being
undistillable, it is useful for entanglement
superactivation~\cite{shor-prl-90-107901}, quantum secret
sharing~\cite{augusiak-pra-73-012318}, or remote information
concentration~\cite{murao-prl-86-352}.  BE states have also given insights into
classical information theory: a classical analog of bound entanglement, called
bound information, exists~\cite{acin-prl-92-107903}.  Recently, a BE state was
simulated with photons~\cite{amselem-natphys-5-748}, as well as in a variant
called pseudo-bound entanglement using nuclear magnetic resonance
(NMR)~\cite{kampermann-pra-81-040304}.

Here, we report the experimentally-observed dynamics of entanglement and
distillability in the neighborhood of a BE state under a partially decohering
environment.  Entanglement and distillability of a multiparty system are
defined with respect to the state \emph{bipartitions}, or abstract splits into
two subsystems.  Our work focuses on a four-party system which can be
bi-partitioned in two ways, either in pairs, or 2:2, and as a single party plus
the rest, or 1:3.  In this case, we call the state of the four particles
\emph{2:2} (\emph{1:3}) \emph{separable} if \emph{every} 2:2 (1:3) bipartition
can be written as a mixture of bipartite states $|\psi_k\rangle$,
\begin{eqnarray*}
\rho=\sum_kp_k|\psi_k\rangle\langle\psi_k|,\;\;\;
|\psi_{k}\rangle=|\eta_{\alpha\beta}^{(k)}\rangle|\eta_{\mu\nu}^{(k)}\rangle\;\;
\left(|\eta_{\alpha}^{(k)}\rangle|\eta_{\beta\mu\nu}^{(k)}\rangle\right),
\end{eqnarray*}
and $\alpha$, $\beta$, $\mu$, and $\nu$ denote the particles.  Otherwise, if
every 2:2 (1:3) bipartition cannot be written as above, we call the state
\emph{2:2} (\emph{1:3}) \emph{entangled}.  Regarding distillability, a state is
\emph{2:2} (\emph{1:3}) distillable if, for \emph{every} 2:2 (1:3) bipartition,
a Bell pair can be distilled and each element of the pair belongs to one
subsystem.  An even stronger distillability property is entanglement
superactivation, which in the case of a four-particle state, it enables five
parties sharing two copies of the state to distill entanglement between the two
parties holding a single particle~\cite{shor-prl-90-107901}.

Our study starts by preparing a 2:2- and 1:3-entangled state, violating a
CHSH-type Bell inequality~\cite{augusiak-pra-73-012318}, and capable of
entanglement superactivation.  As we apply a tunable decohering environment,
the state stops violating the Bell inequality.  Then, only within a region of
further decoherence the entanglement superactivation protocol is successful,
while the 2:2 and 1:3 entanglement is preserved with even more decoherence.
Increasing the decoherence eventually eliminates the entanglement in both
bipartitions at different but finite times.  This finite-time disentanglement
behaviour is also sometimes called environment-induced sudden death of
entanglement~\cite{almeida-science-316-579,yu-science-323-598}.  Since the 2:2
entanglement disappears before the 1:3, we realize a domain, which can be
called BE~\cite{aolita-prl-100-080501}.  Similarly, a recent theoretical study
showed that a four-qubit GHZ state
(Greenberger-Horne-Zeilinger~\cite{horne-ajp-58-1131}) can decay into a BE
state by becoming 2:2 undistillable while slightly 1:3
entangled~\cite{aolita-prl-100-080501}.  Further decoherence eventually makes
the state fully separable, long before the single-particle coherence would
asymptotically disappear.

Our experiment proceeds in three stages: state preparation, exposure to tunable
decoherence, and state characterization.  The goal of the first stage is to
generate a state which decays into the domain of BE states when exposed to a
partially decohering mechanism.  We chose to prepare an initial state
\emph{close} to the Smolin state $\rho_S$~\cite{smolin-pra-63-032306}, a known
four-qubit BE state usually written as a mixture of the Bell states
$\Phi^\pm=(|00\rangle\pm|11\rangle)/\sqrt{2}$ and
$\Psi^\pm=(|01\rangle\pm|10\rangle)/\sqrt{2}$,
\begin{equation}
\rho_S=[|\Phi^+\Phi^+\rangle\langle\cdot|+|\Phi^-\Phi^-\rangle\langle\cdot|+|\Psi^+\Psi^+\rangle\langle\cdot|+|\Psi^-\Psi^-\rangle\langle\cdot|]/4\,,
\end{equation}
where $|0\rangle$ and $|1\rangle$ are the qubit basis states, and we use the
notation $|\chi\rangle\langle\cdot|\equiv|\chi\rangle\langle\chi|$.  The Smolin
state can be prepared by first noticing that it is a mixture of four GHZ-like
states
\begin{eqnarray}
\rho_S=[(|1111\rangle+|0000\rangle)\langle\cdot|+(|1100\rangle+|0011\rangle)\langle\cdot|+\nonumber\\ (|1010\rangle+|0101\rangle)\langle\cdot|+(|1001\rangle+|0110\rangle)\langle\cdot|]/8\;.
\end{eqnarray}
This state can be reached by applying a single-step GHZ-entangling operation to
the mixture
\begin{equation}
\rho=\left[|1111\rangle\langle\cdot|+ |1100\rangle\langle\cdot|+
  |1010\rangle\langle\cdot|+ |1001\rangle\langle\cdot|\right]/4,
\end{equation}
where the operation takes a state of the form $|x_1x_2x_3x_4\rangle$ into
$(|x_1x_2x_3x_4\rangle+|\bar{x}_1\bar{x}_2\bar{x}_3\bar{x}_4\rangle)/\sqrt{2}$
and $\bar{x}_i$ denotes the complement of state $x_i$ of ion $i$ in the
computational basis.  The mixture described by equation~(3), in turn, can be generated
by completely decohering a state in which three out of four particles are
entangled,
\begin{equation}
|\Psi\rangle=\left(|1111\rangle+|1100\rangle+|1010\rangle+|1001\rangle\right)/2\;,
\end{equation}
with the same mechanism as used in the second stage of our study.  Finally,
another GHZ-entangling operation and an NMR-like refocussing technique applied
to the state $|1111\rangle$ generates the state in equation~(4).

In the second stage, the intended rich dynamics was achieved by increasingly
decohering the initial state, as described below (see also Supplementary
Information).  We characterized the state's entanglement and distillability in
the last stage.  A single criterion, the Peres-Horodecki separability
criterion~\cite{peres-prl-77-1413} can prove undistillability while its
extension into a measure, known as negativity~\cite{vidal-pra-65-032314}, can
quantify entanglement.  According to this criterion, if a state is separable
then its partial transposition has no negative eigenvalues (it has a positive
partial transpose, PPT).  On the other hand, it has been shown that PPT states
are undistillable~\cite{horodecki-prl-80-5239}.  Therefore, entangled but
undistillable states, or BE states, can be detected by verifying that
\emph{every} 1:3 bipartition has a negative partial transpose
(entanglement)~\cite{dur-prl-83-3562}, while \emph{every} 2:2 bipartition has a
PPT (undistillability)~\cite{horodecki-prl-80-5239}.  To determine the state's
undistillability properties, we performed a complete tomographic
reconstruction.  Full knowledge of the state enabled us to also check further
separability and distillability properties.  Especially, we designed a novel
algorithm to prove separability of the states, which is a stronger statement
than undistillability (see Supplementary Information).

Our work was performed on a system of four $^{40}$Ca$^+$ ions confined to a
string by a linear Paul trap with axial (radial) vibrational frequencies of
approximately $1.2$ MHz ($4.4$ MHz).  Each ion hosts a qubit on the electronic
Zeeman levels $D_{5/2}(m=-1/2)$, encoding $|0\rangle$, and $S_{1/2}(m=-1/2)$,
encoding $|1\rangle$, determined by a magnetic field of $\approx$ 4 G.  The ion
string was optically pumped to the starting quantum state $|1111\rangle$ after
being Doppler cooled and sideband cooled to the ground state of the axial
center-of-mass (COM) mode~\cite{schmidt-kaler-apblo-77-789}.  The state of the
qubits can be manipulated via (i) collective unitary operations
$U(\theta,\phi)=\exp\left(-i\frac{\theta}{2}S_\phi\right)$, with
$S_\phi=\sum_{k=1}^4\sigma^{(k)}_\phi$,
$\sigma^{(k)}_\phi=\cos(\phi)\sigma^{(k)}_x+\sin(\phi)\sigma^{(k)}_y$, and
$\sigma^{(k)}_n$ a Pauli spin operator acting on the $k$th ion, (ii)
single-qubit light-shift operations
$Z^{(k)}(\theta)=\exp\left(-i\frac{\theta}{2}\sigma^{(k)}_z\right)$, and (iii)
a GHZ-entangling operation known as M\o{}lmer-S\o{}rensen
gate~\cite{molmer-prl-82-1835,sackett-nat-404-256},
$MS(\theta,\phi)=\exp\left(-i\frac{\theta}{4}S_\phi^2\right)$.  We can prepare
four-qubit GHZ states with a fidelity of 96\% and perform collective unitaries
and light-shift operations at a fidelity of 99\%.  These imperfections
determine the proximity of our prepared initial state to the Smolin state.  The
full experimental sequence is shown in Fig.~1.

The partially decohering mechanism indicated in {Figure~1} was
implemented in the four steps shown in {Figure~2:} (i)~hiding the
population in $|0\rangle$ by a full coherent transfer into $S_{1/2}(m=1/2)$;
(ii) transfer of the population in $|1\rangle$ into the superposition
$\sqrt{1-\gamma}|1\rangle+\sqrt{\gamma}|D_{5/2}(m=-5/2)\rangle$; (iii)
quenching of the population in $D_{5/2}(m=-5/2)$ into $P_{3/2}(m=-3/2)$ by
exposure to 854-nm radiation, so that it spontaneously decays to $|1\rangle$;
and finally (iv) restoring the hidden population into $|0\rangle$.  In this
way, a fraction $\gamma$ of the population in $|1\rangle$ irreversibly loses
phase coherence with $|0\rangle$ by tracing over the emitted photon.  In this
case we call this basis-dependent partial loss of coherence \emph{decoherence
  in the $|0\rangle$,$|1\rangle$ basis}.  In our experiment, we decohere the
states in the $|0\rangle\pm|1\rangle$ ($|0\rangle\pm i|1\rangle$) basis by
applying the collective unitary rotations $U(\pi/2,\pi/2)$ ($U(\pi/2,0)$) prior
to the above decoherence, as shown in Fig.~1.  The complete decohering step in
the preparation of the intermediate state in equation~(3) was performed in the
computational basis with $\gamma=1$.

The dynamics of entanglement was explored by varying the amount of decoherence
$\gamma$ to which the initial state was exposed (see Fig.~1).  After being
partially decohered, the density matrices of the prepared states $\rho(\gamma)$
were tomographically reconstructed (see Fig.~3).  Error analysis was performed
via Monte Carlo (MC) simulations over the raw data outcomes of the state
tomography.  The amount of entanglement and signature of undistillability of
the measured states as a function of decoherence $\gamma$ are shown in
Figure~4; the explicit values for the most representative states are quoted in
Table~1.  Figure~4 also indicates other properties of the states determined
independently of the plotted data (full details in Supplementary Information).

The measured initial state ($\gamma=0$) is highly entangled in the 1:3
bipartitions ($N_{1:3}\gg0$) and slightly entangled in the 2:2 ($N_{2:2}>0$);
in addition, it violates a CHSH-type Bell inequality and is capable of
entanglement superactivation.  The properties of the state already change at
$\gamma=0.06$, when the state no longer violates the tested Bell inequality.
The entanglement superactivation protocol is successful in the domain of states
from $\gamma=0$ up to $\gamma=0.12$.  We show strong evidence in the
Supplementary Information that all measured states from $\gamma=0$ to
$\gamma=0.18$ are biseparable.  This means that, although they are entangled
with respect to any fixed 1:3 and 2:2 bipartition, they can be written as a
mixture of separable states, which are separable with respect to
\emph{different} bipartitions.

The passage into bound entanglement occurs at \mbox{$\gamma\approx0.21$}.
While the measured state at $\gamma=0.24$ is 2:2 separable and 1:3 entangled,
the bound entanglement is arguable because a fraction of the MC samples
revealed 2:2 entanglement, thus indicating insufficient statistics.  By
$\gamma=0.32$, the state is now \emph{bona fide} BE, when also all MC samples
are 1:3 entangled, 2:2 undistillable, and even 2:2 separable.
The change into full undistillability is heralded by the state measured at
$\gamma=0.47$ because all eigenvalues of the partial transpose were positive
for every bipartition, as shown in Table~\ref{table:values} and Figure~4.
However, while the measured state is fully separable, known methods failed to
prove the separability of the MC samples.  By $\gamma=0.60$, we achieve full
separability in the measured data and all MC samples.

We also found a BE state by decohering the initial state in \emph{only} the
$|0\rangle\pm|1\rangle$ basis, represented by the state at $\gamma'=0.43$ (see
Table~\ref{table:values}).  We thus controllably realized a passage into bound
entanglement in a simple decohering environment, with statistically significant
entanglement in the 2:2 partitions and positivity of the eigenvalues of the 1:3
partially-transposed states.

In conclusion, we experimentally explored the dynamics of multiparticle
entanglement, separability, and distillability under a tunable decohering
mechanism.  The influence of the environement naturally created a bound
entangled state.  Our investigation on the dynamics of multiparticle
entanglement can be extended to observe bound entanglement on other states such
as decaying GHZ states~\cite{aolita-prl-100-080501} or thermal states of spin
models~\cite{toth-prl-99-250405}.  In addition, recent quantum-computing,
state-engineering, and simulation paradigms driven by dissipative or decohering
environments~\cite{diehl-natphys-4-878,verstraete-nphys-5-633,weimer-nphys-6-382}
can benefit from the environment engineering techniques here demonstrated.

We gratefully acknowledge support by the Austrian Science Fund (FWF), the
European Commission (SCALA, NAMEQUAM, QICS), the Institut f\"ur
Quanteninformation GmbH, and a Marie Curie International Incoming Fellowship
within the 7th European Community Framework Programme.  This material is based
upon work supported in part by IARPA.

Correspondence and requests for materials should be addressed to J.T.B. and O.G.

\begin{table*}[p!]
\caption{Negativity and smallest eigenvalue of the partial transpose of
  representative states.  The parties in the bipartitions are labelled as A, B,
  C, and D.  The state with $\gamma'=0.43$ was prepared by decohering in a
  single step, see text.  Uncertainties in parentheses indicate one standard
  deviation, calculated from propagated statistics in the raw state
  identification events.}
\label{table:values}
\begin{tabular}{l@{\hspace{5mm}}r@{\hspace{3mm}}c@{\hspace{4mm}}c@{\hspace{4mm}}c@{\hspace{5mm}}r@{\hspace{3mm}}c@{\hspace{4mm}}c@{\hspace{4mm}}c@{\hspace{4mm}}c}
%\begin{tabular*}{\textwidth}{@{\extracolsep{\fill}}lrcccrcccc}
%\begin{tabular}{lrcccrcccc}
%\centering{$\gamma$}
& &AB:CD&AC:BD& AD:BC & & A:BCD & B:ACD & C:ACD & D:ABC \\\hline 
$\gamma=0$ & $N_{2:2}=$& 0.033(5) & 0.041(5) & 0.044(5)&
$N_{1:3}=$&0.715(8) & 0.715(8) & 0.715(8) & 0.716(8)\\
$\gamma=0.32$  & 
$\textrm{min}[\textrm{eig}(\rho^{\rm{T}_{2:2}})]=$&0.020(2) & 0.019(2) & 0.022(2)&
$N_{1:3}=$ & 0.035(7) & 0.032(8) & 0.038(8) & 0.045(7)\\
$\gamma=0.47$ & 
$\textrm{min}[\textrm{eig}(\rho^{\rm{T}_{2:2}})]=$& 0.028(3) & 0.029(3) & 0.031(2)&
$\textrm{min}[\textrm{eig}(\rho^{\rm{T}_{1:3}})]=$& 0.019(3) & 0.020(3) & 0.019(3) & 0.018(3)\\\hline
$\gamma'=0.43$ & 
$\textrm{min}[\textrm{eig}(\rho^{\rm{T}_{2:2}})]=$& 0.013(2) & 0.011(2) & 0.013(2)&
$N_{1:3}=$& 0.038(7) & 0.039(7) & 0.042(7) & 0.045(7)\\\hline
\end{tabular}
\end{table*}

%\bibliography{mylibrary}

\begin{figure}
\iffigs\includegraphics{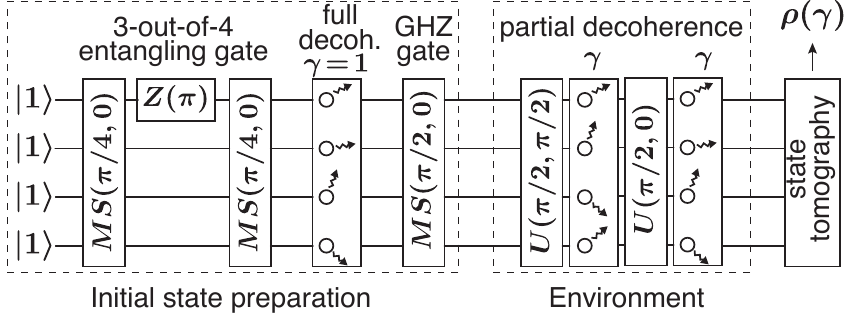}\fi
\label{fig:sequence}
\caption{Experimental sequence for studying the dynamics of multiparticle
  entanglement and distillability induced by decoherence.  The operations
  indicated in the sequence are collective unitary transformations
  $U(\theta,\phi)$, light-shift gates $Z(\theta)$, and entangling gates
  $MS(\theta,\phi)$.}
\end{figure}

\begin{figure}
\iffigs\includegraphics{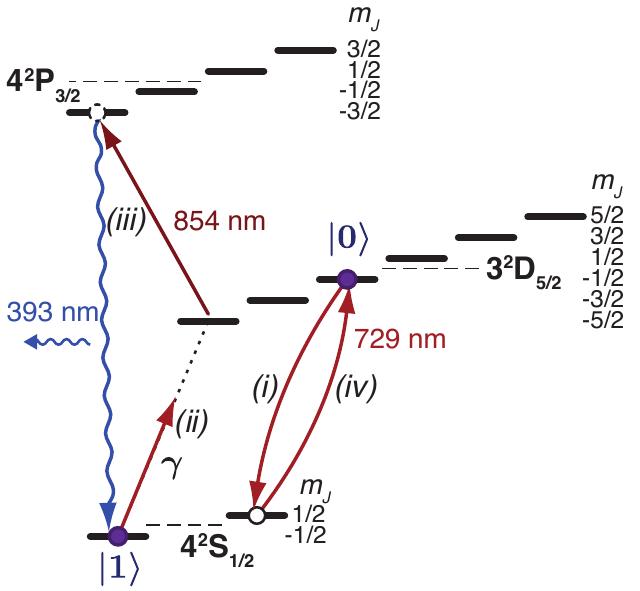}\fi
\label{fig:decoherence}
\caption{Zeeman-split $^{40}$Ca$^+$ levels for the implementation of tunable
  decoherence via entanglement with a spontaneously-decaying photon.  Partial
  decoherence is realized by simultaneously performing on all ions the steps:
  (i) hiding of $|0\rangle$, (ii) partial transfer of $|1\rangle$, (iii)
  quenching and decay of $\sqrt{\gamma}|1\rangle$, and finally (iv) restoring
  $|0\rangle$.  }
\end{figure}

\begin{figure*}
\iffigs\includegraphics{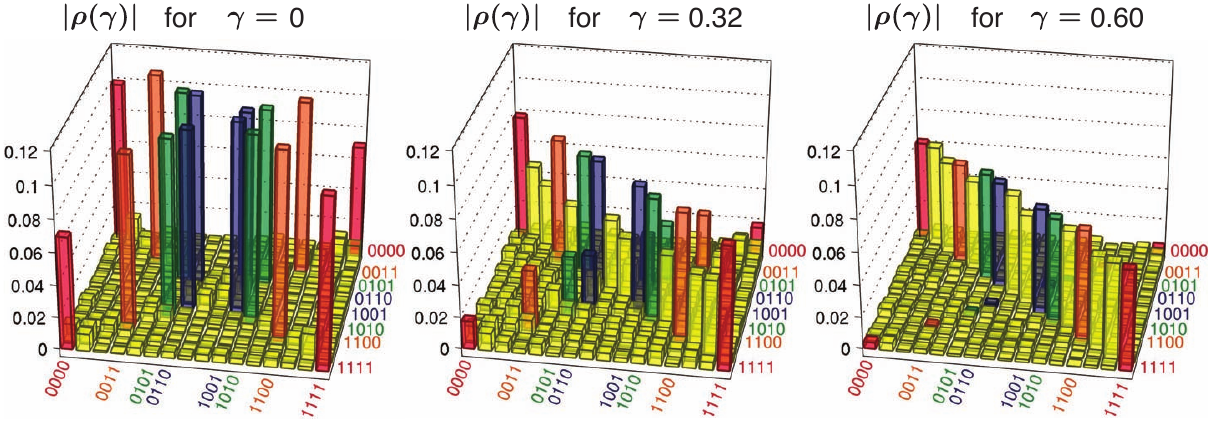}\fi
\label{fig:rhos}
\caption{Density matrices (absolute value) of states which are 2:2 and 1:3
  entangled ($\gamma=0$), bound entangled ($\gamma=0.32$), and fully separable
  ($\gamma=0.60$).  The components of the Smolin state (Eq.~2) are highlighted
  with distinctive colors.  The density matrices of all measured states are
  shown in the Supplementary Information.  }
\end{figure*}

\begin{figure*}
\iffigs\includegraphics{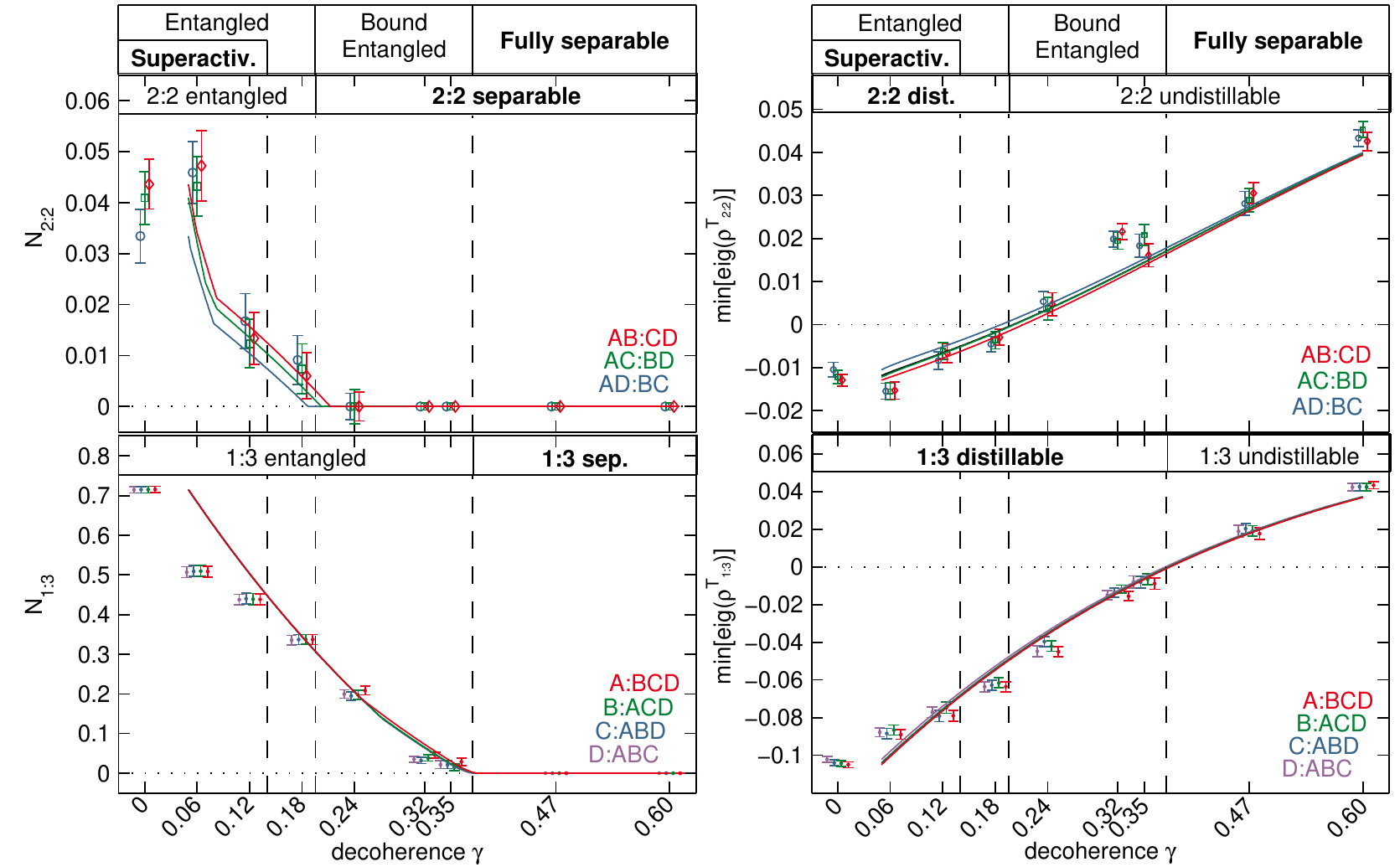}\fi
\label{fig:dataplot}
\caption{Negativity and smallest eigenvalue of the partial transpose for each
  2:2 and 1:3 bipartition of the measured states as a function of decoherence.
  A positive smallest eigenvalue of the partial transpose,
  $\textrm{min}[\textrm{eig}(\rho^{\rm{T}})]>0$, reveals undistillability.
  Bipartitions data are slightly offset horizontally for clarity, but all
  visible groups correspond to the same amount of decoherence indicated by the
  tick marks.  Error bars indicate $\pm$ 1 standard deviation, calculated from
  propagated statistics in the raw state identification events.  The solid
  lines were calculated by decohering the prepared initial state with a 0.05
  offset in $\gamma$ (due to imperfections in the decoherence implementation).
  The properties shown in bold were determined by tests independent of the
  plotted data (see Supplementary Information). }
\end{figure*}

\end{document}

\title{Experimental multiparticle entanglement dynamics induced by
  decoherence\\
SUPPLEMENTARY INFORMATION}

\author{ J. T. Barreiro,$^{1}$ P. Schindler,$^{1}$ O. G\"uhne,$^{2,3}$
  T. Monz,$^{1}$\\ M. Chwalla,$^{1}$ C. F. Roos,$^{1,2}$ M. Hennrich,$^{1}$ and
  R. Blatt$^{1,2}$}

\affiliation{$^{1}$Institut f\"ur Experimentalphysik, Universit\"at Innsbruck,
  Technikerstr. 25, 6020 Innsbruck, Austria\\ 
$^{2}$Institut f\"ur Quantenoptik und Quanteninformation, \\
\"Osterreichische Akademie der Wissenschaften, Technikerstr. 21A, 6020 Innsbruck,
  Austria\\
$^{3}$Institut f\"ur Theoretische Physik,
  Universit\"at Innsbruck, Technikerstr. 25, 6020 Innsbruck,
  Austria}

\maketitle

\tableofcontents
\vspace{5mm}
\noindent\textsf{\textbf{\small List of tables}}\\

\noindent{{\footnotesize
\begin{tabular}[b]{rp{75.3mm}@{\hspace{3mm}}r}
\textbf{I}:&Negativity and smallest eigenvalue of the partial transpose
  for every bipartition of the measured states &{\raggedright5}\\[5mm]
\textbf{II}:&Summary of number of Monte Carlo samples satisfying the
tested properties&5\\[5mm] 
\textbf{III}:&Absolute value of measured density matrices&6\\[3mm]
\textbf{IV}:&Absolute value of measured density matrices after local unitaries
&7\\[5mm] 
\end{tabular}
}}

\section{Proving separability of quantum states}

\subsection{A direct algorithm to prove separability for a bipartite state}

Before explaining our novel algorithm for proving separability of a quantum 
state, let us first define our notations and definitions. We consider a 
bipartite  $N \times M$ system with Hilbert space $\HH=\HH_A \otimes \HH_B.$ 
Any matrix $\vr$ acting on $\HH$ which is Hermitian ($\vr=\vr^\dagger$), 
has no negative eigenvalues ($\vr\geq 0$), and is normalized ($Tr(\vr)=1$) 
is a valid density matrix of some quantum state.

By definition, a state is separable, if it can be written as a convex 
combination of product states,
\be
\vr = \sum_k p_k \ketbra{a_k} \otimes \ketbra{b_k}\,,
\label{sepdef}
\ee
where the $p_k$ are non-negative ($p_k\geq 0$) and normalized 
($\sum_k p_k = 1$); in other words, they form a probability 
distribution. If a state cannot be written as in Eq.~(\ref{sepdef}) 
it is entangled.

For a given experimental $\vr,$ it is not straightforward to prove that it is
separable; finding an explicit decomposition as in Eq.~(\ref{sepdef}) is a
hopeless task. In order to devise a simple algorithm for separability testing,
we use the following two basic facts about separability:

(i) 
Consider a separable state $\vr_{\rm sep}$ and two other 
states $\vr_1$ and $\vr_2$ such that
\be
\vr_1 = p \vr_2 + (1-p) \vr_{\rm sep}
\label{cond1}
\ee for some $p\in [0;1].$ In this situation, if we can prove that $\vr_2$ is
separable, then $\vr_1$ must be separable, too. This follows directly from the
definition in Eq.~(\ref{sepdef}), as this definition implies that the convex
combination of two separable states is separable. Note that if $\vr_2 $ is
entangled this does not mean that $\vr_1$ is entangled.

(ii)
If a state $\vr$ is close to the maximally mixed state 
$\vr_{\rm m}= \tfrac{1}{NM}\eins$, then it is separable. 
Of course, a precise statement of this kind requires 
a specification of a distance. For this, different 
distances have been investigated and bounds
on the distance to $\vr_{\rm m}$ have been obtained 
\cite{braunstein-prl-83-1054,kendon-pra-66-062310,hildebrand-quant-ph-0601201,gurvits-pra-66-062311,gurvits-pra-72-032322,zyczkowski-pra-58-883,vicente-qic-7-624}. 

For our purposes, we will use $Tr(\vr^2)$ as a measure of the mixedness 
and use the fact that if 
\be 
Tr(\vr^2) \leq  \frac{1}{NM-1}
\label{cond2}
\ee
then $\vr$ is separable \cite{gurvits-pra-66-062311}. The maximally mixed state has 
$Tr(\vr_{\rm m}^2)=1/NM$ and states with small $Tr(\vr^2)$ are very 
mixed and close to the maximally mixed state.

Starting from these two facts, the basic idea of our algorithm is to find a 
sequence $\vr_1, \vr_2, \vr_3, ...$ such that  Eq.~(\ref{cond1}) holds for 
all $i,i+1$ and that $Tr(\vr_i^2) \geq Tr(\vr_{i+1}^2)$ holds. This sequence 
will lead to a minimization of  $Tr(\vr_i^2)$ and finally, Eq.~(\ref{cond2})
can be applied to prove separability of $\vr_i$ and hence of $\vr_1.$

In the simplest case, our algorithm can be sketched as follows:

\begin{enumerate}

\item Take the given experimental data $\vr_{\rm exp}$ as $\vr_i$ with $i=1.$

\item  Consider the optimization problem
\be
\max_{\ket{\phi}=\ket{a}\ket{b}} |\bra{\phi} \vr_i \ket{\phi}|
\label{step2}
\ee
and find a product state $\ket{\phi_i}=\ket{a_i}\ket{b_i}$ which has 
a high overlap with $\vr_i.$

Here, the aim is only to find a state with high overlap. In order 
to make the  algorithm work one does not need a certified optimal 
solution of the maximization in Eq.~(\ref{step2}) (see also below).

\item 
Find an $\varepsilon_i$ with $0 \leq \varepsilon_i \leq \varepsilon_{\rm max}$ 
such that
\be
\vr_{i+1} := (1+\varepsilon_i) \vr_i - \varepsilon_i \ketbra{\phi_i}
\label{step3}
\ee
has no negative eigenvalues and that furthermore $Tr(\vr_i^2) \geq Tr(\vr_{i+1}^2)$
holds. Is is natural to choose $\varepsilon_i$ such that $Tr(\vr_{i+1}^2)$ is minimal;
this is, however, not mandatory.

The point here is that if $\ket{\phi_i}$ has a high overlap with 
$\vr_i$, then it will also have a high overlap with the eigenvector 
corresponding to the maximal eigenvalue $\lambda_{\rm max}(\vr_i)$ 
of $\vr_i$. Then, the construction of $\vr_{i+1}$ is such that 
usually $\lambda_{\rm max}(\vr_{i+1}) \leq \lambda_{\rm max}(\vr_{i})$
and, due to the normalization $\lambda_{\rm min}(\vr_{i+1}) \geq \lambda_{\rm min}(\vr_{i})$
holds. Hence, $\vr_{i+1}$ will be closer to the maximally mixed state 
than $\vr_i.$

\item Check, whether $\vr_{i+1}$ fulfills Eq.~(\ref{cond2}). If this is the 
case, then $\vr_{i+1}$ is separable and due to Eq.~(\ref{cond1}) also $\vr_i$ 
and finally $\vr_{\rm exp}$ are separable. Then, the algorithm can terminate.

\item If $\vr_{i+1}$ does not fulfil Eq.~(\ref{cond2}) start again with step 
2 and $i \mapsto i+1$ and iterate further until Eq.~(\ref{cond2}) holds for 
some $i$.

\end{enumerate}

\noindent
This algorithm deserves some comments:

(i) First, it is of course not guaranteed that for a separable input state the
procedure will terminate at some point. Consequently, if it does not terminate
after many steps, one cannot conclude that the state is entangled. It is our
only claim that the algorithm outlined above is a powerful tool in practice.

In Ref.~\cite{navascues-prl-103-160404} an algorithm using semidefinite
programming has been presented which can prove for any non-entangled state the
separability after a finite number of steps. However, the number of steps
required is not known in advance, moreover, the applicability of this technique
to larger Hilbert spaces or multiparticle problems is not clear. Further
algorithms for separability testing have been proposed in
Refs.~\cite{hulpke-jpa-38-5573, spedalieri-pra-76-032318}; there, however, the
practical implementation is still missing.

(ii) Second, note that the algorithm does not require a certified solution of
any non-trivial optimization problem and that it is very robust against
imperfections: Even if the computation of $\ket{\phi_i}$ or $\varepsilon_i$ is
not optimal, this does not affect the conclusion that $\vr_{\rm exp}$ is
separable if Eq.~(\ref{cond2}) holds at some point.  Also, the
$\varepsilon_{\rm max}$ is introduced for practical purposes, chosing a small
$\varepsilon_{\rm max}$ makes the convergence of the algorithm better in
practice.

(iii)
Third, for doing the optimization in Eq.~(\ref{step2}) note that the 
optimal $\ket{\phi}=\ket{a}\ket{b}$ fulfills that $\ket{a}$ is the 
eigenvector corresponding to the maximal eigenvalue of 
$X_A=Tr_B(\vr_i \eins \otimes \ketbra{b})$ and $\ket{b}$ 
is similarly vector corresponding to the maximal eigenvalue of 
$X_B=Tr_A(\vr_i \ketbra{a} \otimes \eins)$. This can be used to 
tackle the maximization iteratively: Starting from a random 
$\ket{a}$ one computes the optimal $\ket{b}$ via $X_B$, 
then with this $\ket{b}$ the optimal $\ket{a'}$, then 
again the optimal $\ket{b'}$ etc. In practice, this converges 
quickly towards the desired solution.

(iv)
Then, one may also check during the iteration whether $\vr_i$ 
violates some of the usual entanglement criteria, e.g.~the 
criterion of the positivity of the partial transpose. If this 
is the case, the iteration will never end and one can directly 
stop it. In this case, however, it is wrong to conclude that 
$\vr_{\rm exp}$ was entangled.

(v)
Moreover, if one wishes to find an explicit separable decomposition 
of $\vr$ as in Eq.~(\ref{sepdef}) the algorithm can also help: 
For some of the states close to the maximally states explicit 
decompositions into product vectors are known \cite{braunstein-prl-83-1054}. 
Then, one can write down a decomposition consisting of this 
decomposition and the $\ket{\phi_i}$ in the iteration.

The algorithm outlined above can be directly implemented with few 
lines of code and performs very well for the states generated in 
the experiment.

\subsection{Proving separability for the multipartite case}

Let us now demonstrate that the ideas from above can also be used to analyze 
multipartite entanglement. Before doing so, we explain some basic notions 
about the entanglement properties of three qubits.

A pure three-qubit state $\ket{\psi}$ is fully separable, if it is of 
the form $\ket{\psi}=\ket{\alpha}\ket{\beta}\ket{\gamma}$ and it is 
biseparable, if there is a grouping of the three parties $A,B,$ and $C$ 
in two partitions (e.g. $AB \vert C$), such that it is separable with respect 
to this partition (e.g. $\ket{\psi}=\ket{\chi}_{AB}\ket{\eta}_C$).
Otherwise, it is genuine tripartite entangled. 
For the special case of three qubits there are further two different 
classes of genuine tripartite entanglement: the GHZ class (represented 
by the GHZ state $\ket{GHZ_3}=(\ket{000}+\ket{111})/\sqrt{2}$) and the 
W class (represented by the W state $\ket{W_3}=(\ket{001}+\ket{010}+\ket{100})/\sqrt{3}$). 
These states represent different classes of entanglement in the sense that a single copy 
of a W-type state cannot be converted locally in a GHZ-type state (and vice versa),
even if the conversion is allowed to work only with a small probability \cite{dur-pra-62-062314}.

As in the bipartite case, one can extend this classification to mixed states 
via convex combinations: A mixed state is fully separable, if it can be written 
as a mixture of fully separable pure states. It is biseparable, if it can be 
written as a mixture of biseparable (and fully separable) states. Otherwise, 
it is genuine multipartite entangled. Here, it is important to note that the 
mixture of biseparable states may contain pure biseparable states which are 
biseperable with respect to different partitions. Furthermore, one can define 
classes of mixed W states and GHZ states \cite{acin-prl-87-040401}.

For a generalization of the separability algorithm it is important that 
similar results  as Eq.~(\ref{cond2}) exist also for multipartite systems
\cite{braunstein-prl-83-1054,kendon-pra-66-062310,gurvits-pra-72-032322,hildebrand-quant-ph-0601201}. For instance, it has been 
shown in Ref.~\cite{hildebrand-quant-ph-0601201} that if a three-qubit state fulfills
\be
Tr(\vr^2) \leq \frac{19}{136} \approx 0.1397,
\label{hildebrandresult}
\ee
then $\vr$ is fully separable. More generally, an N-qubit state with $N \geq 3$ 
is fully separable, if
\be
\label{fullsepbound}
Tr(\vr^2) \leq\frac{1}{2^N-\alpha^2} 
\mbox{ with } 
\alpha^2=\frac{2^N}{\tfrac{17}{2}3^{N-3}+1}
\ee
hold. This follows also from the results of Ref.~\cite{hildebrand-quant-ph-0601201}, where a 
bound on the radius of the separable ball of unnormalized density matrices 
has been given, a rescaling of it  delivers Eq.~(\ref{fullsepbound}).

Given these facts one can now directly write down algorithms to prove full separability
or biseparability. Finding the fully separable (or biseparable) state with the highest 
overlap (see Eq.~(\ref{step2})) can be done as before, by starting from a random fully 
separable (or biseperable) state and then updating it iteratively (see point (iii) above).
Similarly, one can also write an algorithm which can prove that a three-qubit state belongs
to the W class, since the pure W states can be explicitely parameterized \cite{acin-prl-85-1560}.

Again, all these algorithms can be 
easily implemented and they have properties similar to the one discussed before. 

\section{Proving distillability of quantum states}

Let us now explain the criteria that were used for proving distillability of 
the experimentally generated quantum states.

\subsection{Bipartite distillability}
For the one qubit vs. three qubits partitions ($2\times 8$), it is known that 
a quantum state in a $ 2 \times N$ system is distillable, if it has a negative 
partial transpose \cite{dur-pra-61-062313}. Therefore, distillability in this system can 
directly be checked.

For the two qubits vs. two qubits partitions ($4\times 4$) we used the
mathematical definition of distillability, which states that a state $\vr$
on $\HH_A \otimes \HH_B$ is distillable, iff for some $k$ we can find
four states
\bea
\ket{e_i} &\in& \underbrace{\HH_A \otimes...\otimes \HH_A}_{k \;\; {\rm times}} \mbox { for } i \in \{1,2\}
\nonumber
\\
\ket{f_i} &\in& \underbrace{\HH_B \otimes...\otimes \HH_B}_{k \;\; {\rm times}} \mbox { for } i \in \{1,2\}
\eea
such that for $\ket{\psi}=\alpha \ket{e_1}\ket{f_1} +\beta \ket{e_2}\ket{f_2}$ the estimate
\be
\bra{\psi} (\vr^{T_B})^{\otimes k} \ket{\psi} < 0.
\ee
holds \cite{horodecki-prl-80-5239}. For $k=1$ we searched numerically for the
desired $\ket{\psi}$, and if this can be found, the state must be distillable.

\subsection{Entanglement superactivation}

Entanglement superactivation is a quantum information processing task, 
which
can work also for the Smolin state as a bound entangled state. Consider two 
copies of a four-qubit state $\vr$ which is distributed among five parties
as 
\begin{equation}
\vr_{\rm total} = \vr^{ABCD} \otimes \vr^{ABCE},
\end{equation}
that is, the parties $A,B,$ and $C$ hold two qubits each, while the 
parties $D$ and $E$ have only one. In this situation, it was shown 
that if $\vr$ is a Smolin state, then the five parties can create a 
Bell state between $D$ and $E$ by local operations and classical communication
only \cite{shor-prl-90-107901}. The protocol uses a sequence of teleportations.

For the experimental data, we have first applied appropriate local unitary
rotations, then this protocol and finally checked with the PPT criterion
whether the resulting state between $D$ and $E$ is entangled. We also tested
this for arbitrary permutations of the four qubits, since the experimental data
are, in contrast to the ideal Smolin state, not permutationally invariant.

\section{CHSH-type Bell inequality}

We consider the scenario where each of the 4 parties, labeled $j$
($j=1,2,3,4$), can choose between two observables $(O_{j}^{{k}_{j}})$,
$k_{j}=1,2$.  The CHSH-type Bell inequalities considered here have the
form~\cite{werner-pra-64-032112,zukowski-prl-88-210401,augusiak-pra-74-010305}:
\begin{equation}\label{nBella}
|E(1,1,1,1)+E(1,1,1,2)+E(2,2,2,1)-E(2,2,2,2)| \le 2\,,
\end{equation}
where the correlation function $E$ for the measured quantum state $\rho$ is an
average calculated as follows:
\begin{equation}\label{3.3}
E(k_{1},k_{2},k_{3},k_{N})(\varrho)=\textrm{Tr}\left[\rho
O_{k_{1}}^{(1)} \otimes O_{k_{1}}^{(2)} \otimes O_{k_{1}}^{(3)} \otimes O_{k_{N}}^{(4)}\right].
\end{equation}
For the measured states we maximize the left-hand side of the inequality in
Eq.~\ref{nBella}, also known as Bell parameter, by varying the observable
directions.  For the state with $\gamma=0$, this Bell parameter is 2.21(2),
while for the state with $\gamma=0.06$, the parameter is already 1.47(3).

\section{Biseparability of the states with $\gamma=0,0.06,0.12,$ or $0.18.$}

Applying the separability algorithm to the measured states for these values of
$\gamma$ does not lead to a proof that these states are biseparable. This is
due to the fact that the experimental states have some eigenvalues which are
practically zero, making the third step of the separability algorithm
(Eq.~\ref{step3}) difficult. However, if one takes the state with $\gamma=0$
and adds 2\% of white noise
\begin{equation}
\vr_{\rm noise} = 0.98 \vr(\gamma=0) + 0.02 \frac{\openone}{16}
\end{equation}
the algorithm proves that this state $\vr_{\rm noise}$ is biseparable.  It
should be noted that the statistical fluctuations of the diagonal elements of
$\vr(\gamma=0)$ are larger than the amount of noise added.  Moreover, if one
considers the state consisting of the mixture of all Monte Carlo sample states,
the algorithm proves that this state is biseparable.  This gives strong
evidence for the biseparability of the experimentally generated states.

\section{Model of imperfections reveals the observed dynamics}

Several imperfections affect the experimental setup, but to briefly demonstrate
that the experimentally observed entanglement dynamics agrees with
straightforward calculations, here we choose to only assume imperfect
M\o{}lmer-S\o{}rensen operation times (angle $\theta$ in $MS(\theta,\phi)$, see
report).  We consider longer operation times by a small fraction,
$\varepsilon_1$ for the 3-out-of-4-entangling operation,
$MS((1+\varepsilon_1)\pi/4,0)$, and $\varepsilon_2$ for the GHZ-entangling
operation, $MS((1+\varepsilon_2)\pi/2,0)$.  Using the preparation sequence
shown in Figure~1 of our report and assuming other operations are perfectly
realized, we calculate an expected initial state.  Upon decohering this
imperfect initial state in two steps, as described in the report, we observe
the dynamics of entanglement resembling our measurements, as illustrated here
in Figure~1.

\begin{figure}[!h]
\centerline{\includegraphics[width=86mm]{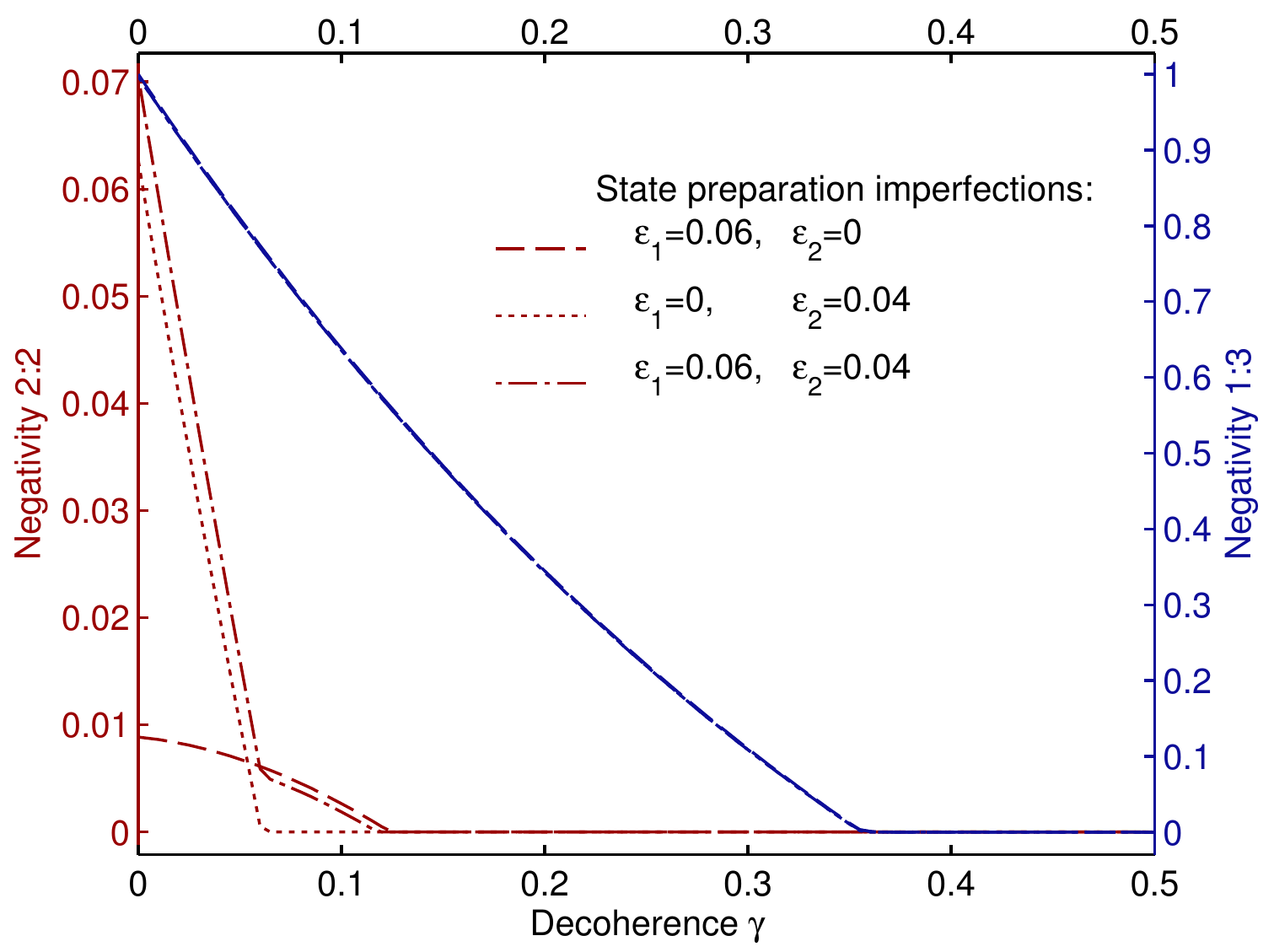}}
\label{fig:modeldecay}
\caption{Negativity for the 2:2 bipartition (left axis) and 1:3 bipartition
  (right axis) as a function of partial decoherence for the shown imperfections
  (see text).  The imperfections have no effect on the dynamics of $N_{1:3}$.
  The negativity is the same for all permutations of each bipartition due to
  the symmetry of the initial state.}  \nonumber
\end{figure}

\section{Quantum state reconstruction and statistics}

Physical states were tomographically reconstructed via maximum likelihood
estimation (MLE)~\cite{jezek-pra-68-012305}.  On the four qubits, we performed
$3\times3\times3\times3=81$ measurements by iterating the measurement of the
expectation values of $\sigma_x$, $\sigma_y$, and $\sigma_z$ on each qubit.
For each state, the number of copies used per state are shown in
Table~\ref{table:peresvalues}.  Tomographic data were acquired in less than 5.5
hours for each state.  Full tomographic sets of 200-250 copies per measurement
were acquired iteratively to keep track of potential drifts in the experimental
setup.

Error analysis was calculated via Monte Carlo simulations over the
multinomially distributed measurement outcomes of the state tomography.  For
each state, 200 Monte Carlo samples were generated and reconstructed via MLE.

%\bibliography{mylibrary}

\clearpage

\widetext

\noindent
\begin{table*}[!h]
\caption{Negativity and smallest eigenvalue of the partial transpose of
  measured states for every bipartition.}
\label{table:peresvalues}
{{\small
\begin{tabular}{|c|c|cccc|ccc|cc|c|}\cline{3-11}
\multicolumn{2}{c}{}&\multicolumn{4}{|c|}{2:2 \& 1:3 entangled}&\multicolumn{3}{|c|}{Bound
    Entangled}&\multicolumn{2}{|c|}{Fully
  Separable}&single\footnote{Bound-entangled state decohered in a single step,
  see report.}\\
\multicolumn{2}{r|}{$\gamma=$}  & 0.00 & 0.06 & 0.12 & 0.18 & 0.24 & 0.32 &
0.35 & 0.47 & 0.60 & 0.43\\\hline
\multirow{3}{*}{$N_{2:2}$}
& AB:CD & 0.033(5)  & 0.046(6)  & 0.017(5)  & 0.009(5)  & 0.000(3)  & 0  & 0  &
0  & 0 & 0\\
& AC:BD & 0.041(5)  & 0.043(6)  & 0.012(5)  & 0.007(5)  & 0.000(3)  & 0  & 0  &
0  & 0 & 0\\
& AD:BC & 0.044(5)  & 0.047(7)  & 0.013(5)  & 0.006(5)  & 0.000(3)  & 0  & 0  &
0  & 0 & 0\\\hline
\multirow{3}{*}{$\mathrm{min}(\mathrm{eig}(\rho^{\mathrm{T}_{2:2}}))$}
& AB:CD & -0.011(2) & -0.016(2) & -0.008(2) & -0.005(2) & 0.005(2) & 0.020(2) &
0.018(3) & 0.028(3) & 0.043(2) & 0.013(2)\\
& AC:BD & -0.012(2) & -0.016(2) & -0.006(2) & -0.004(2) & 0.004(3) & 0.019(2) &
0.021(2) & 0.029(3) & 0.045(2) & 0.011(2)\\
& AD:BC & -0.013(1) & -0.015(2) & -0.007(2) & -0.003(2) & 0.005(3) & 0.022(2) &
0.016(3) & 0.031(2) & 0.043(2) & 0.013(2)\\\hline
\multirow{4}{*}{$N_{1:3}$}
& A:BCD & 0.715(8)  & 0.507(14)  & 0.438(14)  & 0.335(13)  & 0.199(11)  &
0.035(7)  & 0.021(10)  & 0  & 0 & 0.038(7)\\
& B:ACD & 0.715(8)  & 0.509(14)  & 0.440(14)  & 0.337(12)  & 0.195(11)  &
0.032(8)  & 0.022(10)  & 0  & 0 & 0.039(7)\\
& C:ABD & 0.715(8)  & 0.510(14)  & 0.439(14)  & 0.337(12)  & 0.197(11)  &
0.038(8)  & 0.015(9)  & 0  & 0 & 0.042(7)\\
& D:ABC & 0.716(8)  & 0.509(14)  & 0.438(14)  & 0.337(12)  & 0.208(11)  &
0.045(7)  & 0.028(10)  & 0  & 0 & 0.045(7)\\\hline
\multirow{3}{*}{$\mathrm{min}(\mathrm{eig}(\rho^{\mathrm{T}_{2:2}}))$}
& A:BCD & -0.102(2)  & -0.088(2)  & -0.077(3)  & -0.064(3)  & -0.045(3)  &
-0.015(2) & -0.008(3)  & 0.019(3)  & 0.042(2) & -0.011(2)\\
& B:ACD & -0.104(2)  & -0.088(3)  & -0.079(3)  & -0.063(3)  & -0.040(3)  &
-0.014(2) & -0.008(3)  & 0.020(3)  & 0.042(2)  & -0.015(2)\\
& C:ABD & -0.104(1)  & -0.087(3)  & -0.075(3)  & -0.062(3)  & -0.042(3)  &
-0.012(2) & -0.007(3)  & 0.019(3)  & 0.042(2) & -0.014(2)\\
& D:ABC & -0.105(2)  & -0.089(2)  & -0.079(3)  & -0.064(3)  & -0.045(3)  &
-0.016(2) & -0.009(3)  & 0.018(3)  & 0.043(2) & -0.014(2)\\\hline
\multicolumn{2}{|c|}{(state copies)/(81 meas.)} & 5000 & 3000 & 3000 & 3250 & 3750 &
6000 & 3100 & 3000 & 4000 & 6000\\\hline
\end{tabular}
}}
\end{table*}

\normalsize

\noindent
\begin{table*}[!h]
\caption{Summary of Monte Carlo samples satisfying the tested properties.}
\label{table:tests}
\begin{tabular*}{\textwidth}{@{\extracolsep{\fill}}|l|c|c|c|c|c|c|c|c|c|c|}\hline
Decoherence $\gamma=$&0& 0.06 & 0.12 & 0.18 & 0.24 & 0.32 & 0.35 & 0.47 &
0.60 & single\\\hline 
\hline
2:2 entangled & 200& 200&
200& 200&41\footnote{These 41 states had a non-positive partial transpose for all
  permutations of the 2:2 bipartition. }
&0
&0
&0 &0&1\\
2:2 distillable &200&200&194&172&2\footnote{44 states had a positive partial
  transpose on every permutation of the 2:2 bipartition (undistillable) and were also 2:2 separable.}&0&0&0&0&0\\
1:3 entangled & 200& 200& 200& 200& 200& 200& 200 & 0& 0&200\\ 
1:3 distillable &200&200&200&200&200&200&200&0&0&200\\
superactivatable & 200 & 200 & 200 &0&0&0&0&0&0&0\\
biseparable&200\footnote{Strong evidence, discussed in
  text.\label{fn:se}}&200\footref{fn:se}&200\footref{fn:se}&200\footref{fn:se}&200&200&200&200&200&200\\ 
fully
separable&0&0&0&0&0&0&0&0
& 200 & 0\\\hline
\end{tabular*}
\end{table*}

\noindent
\begin{table*}[!h]
\caption{Absolute value of measured density matrices.}
\begin{tabular*}{.90\textwidth}{@{\extracolsep{\fill}}ccc}
$|(\rho(\gamma)|$ for $\gamma=0$   &
$|(\rho(\gamma)|$ for $\gamma=0.06$&
$|(\rho(\gamma)|$ for $\gamma=0.12$\\
\includegraphics[width=5cm]{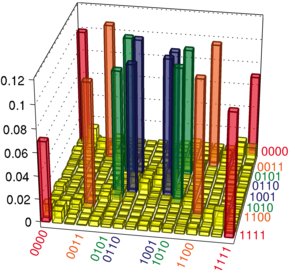}&
\includegraphics[width=5cm]{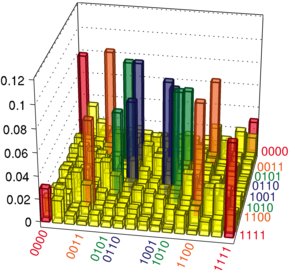}&
\includegraphics[width=5cm]{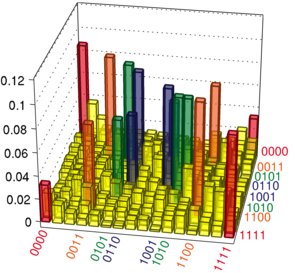}\\
$|(\rho(\gamma)|$ for $\gamma=0.18$&
$|(\rho(\gamma)|$ for $\gamma=0.24$&
$|(\rho(\gamma)|$ for $\gamma=0.32$\\
\includegraphics[width=5cm]{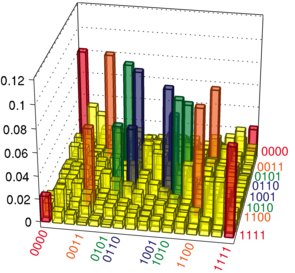}&
\includegraphics[width=5cm]{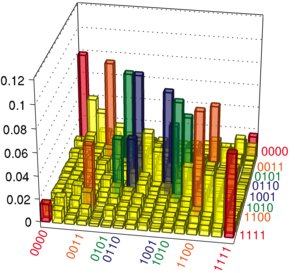}&
\includegraphics[width=5cm]{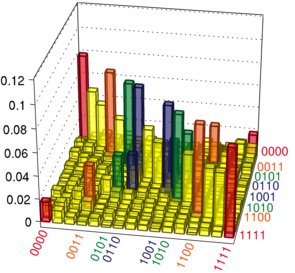}\\
$|(\rho(\gamma)|$ for $\gamma=0.35$&
$|(\rho(\gamma)|$ for $\gamma=0.47$&
$|(\rho(\gamma)|$ for $\gamma=0.60$\\
\includegraphics[width=5cm]{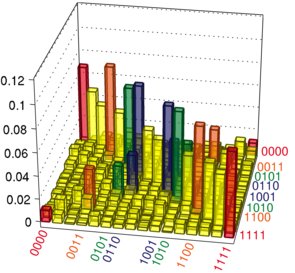}&
\includegraphics[width=5cm]{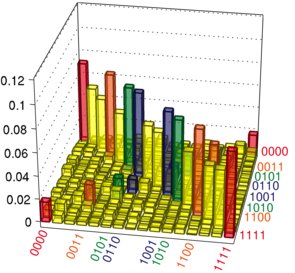}&
\includegraphics[width=5cm]{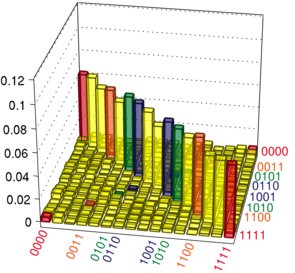}\\
&$|(\rho(\gamma)|$ for $\gamma'=0.43$\\
&\includegraphics[width=5cm]{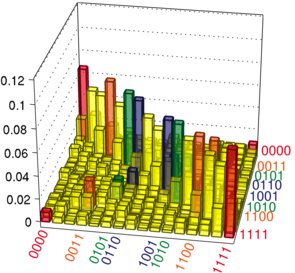}\\
\end{tabular*}
\end{table*}

\noindent
\begin{table*}[!h]
\caption{Absolute value of density matrices $\varrho(\gamma)$ calculated by
  applying local unitaries to the measured states $\rho(\gamma)$ which maximize
  the amplitudes of their real part.  This transformation facilitates
  visualizing the Smolin components of the states.}
\begin{tabular*}{.90\textwidth}{@{\extracolsep{\fill}}ccc}
$|(\varrho(\gamma)|$ for $\gamma=0$   &
$|(\varrho(\gamma)|$ for $\gamma=0.06$&
$|(\varrho(\gamma)|$ for $\gamma=0.12$\\
\includegraphics[width=5cm]{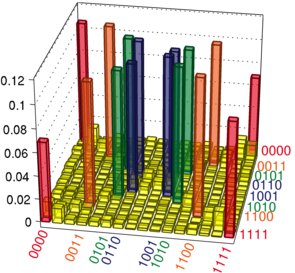}&
\includegraphics[width=5cm]{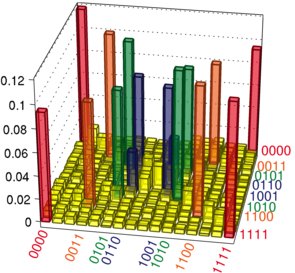}&
\includegraphics[width=5cm]{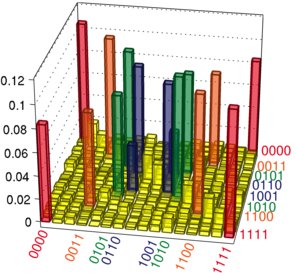}\\
$|(\varrho(\gamma)|$ for $\gamma=0.18$&
$|(\varrho(\gamma)|$ for $\gamma=0.24$&
$|(\varrho(\gamma)|$ for $\gamma=0.32$\\
\includegraphics[width=5cm]{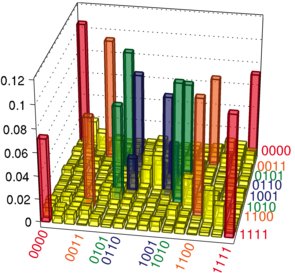}&
\includegraphics[width=5cm]{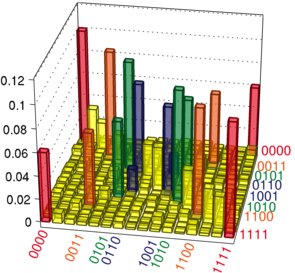}&
\includegraphics[width=5cm]{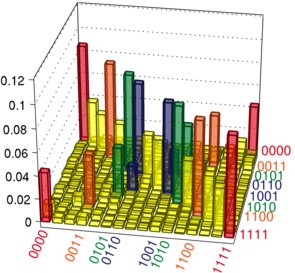}\\
$|(\varrho(\gamma)|$ for $\gamma=0.35$&
$|(\varrho(\gamma)|$ for $\gamma=0.47$&
$|(\varrho(\gamma)|$ for $\gamma=0.60$\\
\includegraphics[width=5cm]{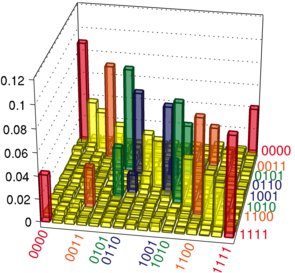}&
\includegraphics[width=5cm]{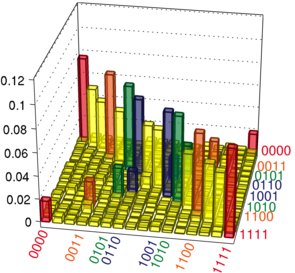}&
\includegraphics[width=5cm]{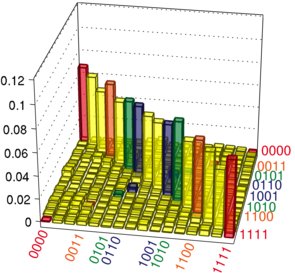}\\
&$|(\varrho(\gamma)|$ for $\gamma'=0.43$\\
&\includegraphics[width=5cm]{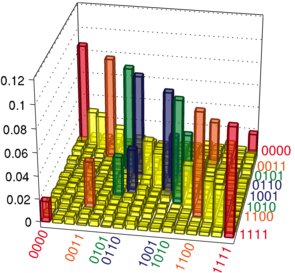}\\
\end{tabular*}
\end{table*}